\documentclass[12pt]{iopart}

\usepackage{iopams}

\usepackage{graphicx}
\usepackage{dcolumn}

\begin{document}

\newcommand{\dmskcomment}[1]{\marginpar{$\spadesuit$}}
\newcommand{\ftot}{{F_{\mbox{\scriptsize tot}}}}
\newcommand{\gammamax}{\gamma_{\mbox{\scriptsize max}}}
\newcommand{\hamispin}{\mathcal{H}_{\mbox{\scriptsize{spin}}}}
\newcommand{\hami}{\mathcal{H}}
\newcommand{\kmax}{k_{\mbox{\scriptsize{max}}}}
\newcommand{\mone}{M^{(1)}}
\newcommand{\mtwo}{M^{(2)}}
\newcommand{\hamiint}{\mathcal{H}_{\mbox{\scriptsize{int}}}}
\newcommand{\mtildeone}{\tilde{M}^{(1)}}
\newcommand{\mtildetwo}{\tilde{M}^{(2)}}

\title{Spin squeezing of high-spin, spatially extended quantum fields}

\author{Jay D. Sau$^{1,2}$}
\author{S. R. Leslie$^{1}$}
\author{Marvin L. Cohen$^{1,3}$}
\author{D. M. Stamper-Kurn$^{1,3}$}
\ead{dmsk@berkeley.edu}  
\address{
    $^1$Department of Physics, University of California, Berkeley, California 94720, USA \\
    $^2$Condensed Matter Theory Center and Joint Quantum Institute, Department of Physics,
University of Maryland, College Park, Maryland 20742-4111, USA\\
    $^3$Materials Sciences Division, Lawrence Berkeley National Laboratory, Berkeley, California 94720}
\date{\today}%

\begin{abstract}
Investigations of spin squeezing in ensembles of quantum particles
have been limited primarily to a subspace of spin fluctuations and a
single spatial mode in high-spin and spatially extended ensembles.
Here, we show that a wider range of spin-squeezing is attainable in
ensembles of high-spin atoms, characterized by sub-quantum-limited
fluctuations in several independent planes of spin-fluctuation
observables. Further, considering the quantum dynamics of an $f=1$
ferromagnetic spinor Bose-Einstein condensate, we demonstrate
theoretically that a high degree of spin squeezing is attained in
multiple spatial modes of a spatially extended quantum field, and
that such squeezing can be extracted from spatially resolved
measurements of magnetization and nematicity, i.e.\ the vector and
quadrupole magnetic moments, of the quantum gas. Taking into account
several experimental limitations, we predict that the variance of
the atomic magnetization and nematicity may be reduced as far as 20
dB below the standard quantum limits.
\end{abstract}

\maketitle

Spin squeezing describes a form of entanglement in a many-spin
system for which the variance in measurements of particular
collective spin variables is smaller than the minimum variance
achievable in uncorrelated systems \cite{wine92squeeze,kita93}.
Aside from representing a theoretically tractable form of
entanglement in many-body quantum systems, spin squeezing also
promises to improve measurement precision in a variety of
applications such as chronometry, magnetometry, and atom
interferometry.  The recent achievement of metrologically beneficial
pseudo-spin squeezing in cold atomic gases, on the
Zeeman-insensitive hyperfine transitions utilized in atomic clocks,
is among the first realizations of this promise
\cite{schl10states,appe09}.

The conventional form of spin squeezing involves the collective
vector spin operator, $\mathbf{F}$, which is the equal-weight sum of
the vector spin operators for all of the ensemble's constituent
particles. Given an average ensemble spin oriented along the
$\hat{z}$ axis, the measurement variance of orthogonal spin
projections obey an uncertainty relation, $\langle (\Delta
F_x)^2\rangle \langle (\Delta F_y)^2 \rangle \geq \langle F_z
\rangle^2/4$. Ensemble states of uncorrelated particles are limited
by the standard quantum limited variance, $\langle (\Delta
F_\perp)^2 \rangle_{SQL} = |\langle F_z \rangle|/2$, achieved by
partitioning the minimum measurement uncertainties equally between
the two vector spin projections.

This form of \emph{vector operator} spin squeezing describes present
experiments not only on pseudo-spin 1/2 atomic gases, i.e.\ where
each atom is restricted to just two internal
\cite{schl10states,appe09} or external \cite{este08squeeze} states,
but also on gases of higher-spin atoms (e.g.\ the $f = 4$
hyperfine-state manifold of Cs \cite{hald99}). However, the
description of higher-spin atoms by their vector spin alone is
incomplete.  On the one hand, light-atom interactions that affect
higher-order spin moments may impair efforts to create vector-spin
squeezing in high-spin atomic gases \cite{kupr05,stoc06tensor}. On
the other hand, the deliberate addressing and measurement of
higher-order spin moments \cite{yash03higher} offers benefits to
certain metrological applications, e.g.\ eliminating directional
dependencies of atomic magnetometers \cite{acos08hexa}.

Here, we consider theoretically the characterization and preparation
of spin-squeezed states in $f=1$ spinor Bose gases. We go beyond
existing treatments of spin squeezing by evaluating such squeezing
in an ensemble  that is explicitly higher-dimensional not only in
its spin degrees of freedom, but also in its spatial degrees of
freedom. Our investigation is motivated by two experimental results:
the controlled amplification of spin noise in $^{87}$Rb quantum
fluids \cite{sadl06symm,lesl09amp,klem09multi}, and the use of a
spinor Bose-Einstein condensate as a spatially resolving
magnetometer \cite{veng07mag}. As we discuss herein, these two
results, respectively, provide a complete means for inducing
squeezing of both the spin vector and quadrupole moments of the
$f=1$ quantum gas, and also a means of applying the subsequent
spatially extended non-classical quantum fluid toward spatially
resolved magnetometry with sub shot-noise sensitivity.

\section{Single-mode squeezing for arbitrary spin}
\label{sec:singlemode}

We begin by introducing a generalized treatment of spin squeezing
that applies equally to ensembles of spin-1/2 and also of
higher-spin particles.  Such squeezing will be defined with respect
to a coherent spin state \cite{kita93}, in which all particles are
prepared in an identical single-particle spin-$f$ state
$|\psi\rangle$.  Consider an orthogonal basis of the single-particle
spin space that includes the  state $|\psi\rangle$ and also the $2
f$ states $|\xi_a\rangle$. We now define many-body observables
$M^{(1)}_a = \sum_\alpha \left(|\xi_a\rangle \langle \psi | +
|\psi\rangle \langle \xi_a | \right)_\alpha$ and $M^{(2)}_a =
\sum_\alpha i \left(|\xi_a\rangle \langle \psi | - |\psi\rangle
\langle \xi_a | \right)_\alpha$ where the summation index $\alpha$
is taken over all particles in the $N$-particle ensemble. These spin
fluctuation operators describe the subspace of spin excitations
orthogonal to the coherent spin state. With respect to this
uncorrelated state, these observables satisfy the commutation
relations $\langle [ M^{(1)}_a, M^{(2)}_b]\rangle = 2 i N
\delta_{a,b}$. Thus, measurements of each pair $M^{(1,2)}_a$
independently obey an uncertainty relation $\langle ( \Delta
M^{(1)}_a)^2 \rangle \langle ( \Delta M^{(2)}_a )^2\rangle \geq
N^2$. Following the treatment of squeezing for vector spin
operators, we identify one mode of spin fluctuations, labeled by
$a$, to be spin squeezed when the measurement variance for one
quadrature of the $M_a^{(1)}$-$M_a^{(2)}$ plane is below the
standard quantum limit, $N$.

Applying this treatment to an ensemble of spin-1/2 particles, we
recall that any coherent spin state can be written as the $m=+1/2$
eigenstate of a projection of the dimensionless vector spin.  Taking
that projection to lie along $\hat{z}$, we identify the one mode of
spin fluctuation operators as the Pauli operators $\sigma_x$ and
$\sigma_y$. More generally, for an ensemble of spin-$f$ particles
prepared in the $|\psi \rangle  = |m_z = f\rangle$ eigenstate of the
$F_z$ spin operator, the spin fluctuation operators defined by the
state $|\xi_1\rangle = |m_z = f-1\rangle$ act on $|\psi\rangle$ as
$\sqrt{2/f}$ times the spin vector operators $F_x$ and $F_y$. For
this one mode of spin-fluctuation operators, the commutation
relation matches that of the spin-vector operators.  Thus, we
recover the conventional description of vector-spin squeezing as
pertaining to just one mode of spin fluctuations atop a maximum-spin
coherent spin state.

To illustrate further the possibility of squeezing in several spin
fluctuation modes, we consider spin squeezing of spin-1 particles
prepared initially in the $|m_z = 0\rangle$ state. Our treatment is
facilitated by working in a polar state basis, where
$|\phi_e\rangle$ is the zero-eigenvalue state of ${\bf{F}} \cdot
{\bf{e}}$ \cite{note:polar}. The pair of fluctuation operators
defined according to the state $|\phi_x\rangle$ ($|\phi_y\rangle$)
are identified as $M^{(1)} = -N_{xz}$ ($-N_{yz}$) and  $M^{(2)} =
+F_{y}$ ($-F_{x}$), where the former is a component of the
quadrupole moment tensor \cite{note:quadrupole}. The coherent spin
state $|m_z = 0 \rangle$ is thus regarded as the uncorrelated vacuum
state for independent spin fluctuations in the $F_x$-$N_{yz}$ and
the $F_y$-$N_{xz}$ planes, and correlated states may be spin
squeezed with respect to vector-spin or quadrupole-spin components,
or linear combinations thereof. Finally, we note that this
unconventional form of spin squeezing, about the polar
$|m_z=0\rangle$ state, may be mapped directly onto the more
conventional vector-spin squeezing discussed above by a unitary
transformation that rotates the polar basis states onto the proper
$F_z$ eigenstates; experimentally, such generic unitary
transformations upon alkali-atom hyperfine spin states are produced
by rf-pulses and quadratic Zeeman energy shifts \cite{gior03}.

\section{Spin squeezing in a spatially extended quantum field}
\label{sec:spatial}

We now consider such multi-spin-mode squeezing in a spatially
extended, ferromagnetic $f=1$ gaseous Bose-Einstein condensate
where, as we show below, such squeezing is produced naturally by
spin-dependent interactions \cite{note:necessary}.   As in recent
experiments \cite{sadl06symm,lesl09amp}, these condensed atoms (at
rest) are prepared initially in the common spin state $|m_z =
0\rangle$.  Within the volume occupied by the condensate, we define
position-space spin-fluctuation measurement densities
\begin{eqnarray}
M^{(1)}_a(\mathbf{r}) & = & \left( \phi^\dagger_a(\mathbf{r}) \phi_z(\mathbf{r}) +  \phi^\dagger_z(\mathbf{r}) \phi_a(\mathbf{r}) \right) \\
M^{(2)}_a(\mathbf{r}) & = & i \left( \phi^\dagger_a(\mathbf{r})
\phi_z(\mathbf{r}) -  \phi^\dagger_z(\mathbf{r}) \phi_a(\mathbf{r})
\right)
\end{eqnarray}
where $\phi_e(\mathbf{r})$ is now the position-space Bose field
operator for the polar state $|\phi_e\rangle$, and the index $a \in
\{x,y\}$ runs over the transverse spin polarizations.  These
observables represent components of the local nematicity (rank-2
moments of the atomic spin) and magnetization (rank-1 moments of the
atomic spin) of the quantum gas, and obey the commutation relation $
\left\langle \left[ M^{(1)}_a(\mathbf{r}),
M^{(2)}_b(\mathbf{r}^\prime) \right] \right\rangle = 2 i \,
n({\mathbf{r}}) \, \delta^3(\mathbf{r} - \mathbf{r}^\prime)
\delta_{a,b} $ where $n({\mathbf{r}}) = \langle
\phi^\dagger_z(\mathbf{r}) \phi_z(\mathbf{r}) \rangle$, and
expectation values are taken with respect to the polar condensate.

These measurement density operators are used to construct operators
corresponding to spatially resolved measurements of particular spin
fluctuations of the quantum field.  Defining the real measurement
mode functions $\{A(\mathbf{r}), B(\mathbf{r}), C(\mathbf{r}),
D(\mathbf{r})\}$, we may define mode measurement operators as
follows:
\begin{equation}
\left(\begin{array}{c} \mone_{A, B, C, D}
\\ \mtwo_{A, B, C, D} \end{array} \right) = \int d^3\mathbf{r} \, \left( \begin{array}{c c} A(\mathbf{r}) &  B(\mathbf{r}) \\
C(\mathbf{r})& D(\mathbf{r}) \end{array} \right)
\left(\begin{array}{c} \mone(\mathbf{r})
\\ \mtwo(\mathbf{r}) \end{array} \right)
\end{equation}
where the common polarization index is omitted for clarity.  In this
definition for $\mone_{A,B,C,D}$ (and similarly for
$\mtwo_{A,B,C,D}$), the local measurement weight is given as
$\sqrt{A^2 + B^2}$ and the locally measured quadrature is defined by
the angle $\tan^{-1}(B/A)$. The commutation relation
\begin{equation}
\left\langle [\mone_{A, B, C, D} , \mtwo_{A, B, C, D} ]
\right\rangle = 2 i\, \int d^3\mathbf{r} \, \left( A D- B C\right)
n(\mathbf{r})
\end{equation}
is non zero to the extent that the two measurement operators probe
non-parallel spin-fluctuation quadratures on a common population of
atoms.

We see that our present treatment subsumes the commonly discussed
single-mode description of spin squeezing in atomic ensembles.  That
is, the equal-weight sum of atomic spin operators is measured by
spatial-mode operators defined with the uniform functions $A\!=
\!D\!=\! 1$ and $B\!=\! C\!=\! 0$. However, our treatment easily
allows for a multi-mode description of spin squeezing and collective
spin measurements of a spatially extended quantum field. For
example, spin squeezing in a spatially resolved measurement of the
atomic spin, e.g.\ in the spinor-gas magnetometer of Ref.\
\cite{veng07mag}, may be treated by defining measurement mode
functions corresponding to every resolved pixel in the
magnetization-sensitive image.  Our treatment applies also to
optical measurements sensitive to spatially varying linear
combinations of magnetization and nematicity, which are achieved by
measuring different components of the linear optical susceptibility
tensor \cite{caru04imag}.

\section{Generation of spin squeezing in a ferromagnetic $f=1$
Bose-Einstein condensate}

Now, let us consider the evolution of the paramagnetic condensate
under the quantum-coherent spin dynamics produced by the
spin-dependent contact interactions between atoms
\cite{lama07quench,mias08}.  As in previous experiments, we consider
the condensate to be prepared in the uncorrelated polar state and
then allowed to evolve freely at a fixed value of the quadratic
Zeeman shift $q$, which gives the difference between the average
energy of the $|m_z = \pm 1\rangle$ Zeeman sublevels and that of the
$|m_z=0\rangle$ state. Within the Bogoliubov approximation, the
spin-dependent Hamiltonian may be written as follows:
\begin{equation}
\hamispin = \sum_a \int d^3\mathbf{r}\, \left[\phi_a^\dagger \left(
\hami_0 + \hamiint \right)\phi_a + \frac{\hamiint}{2} \left( \left.
\phi^{\dagger}_a\right.^2 + \phi_a^2\right)\right]
\label{eq:hspinbogo}
\end{equation}
where $\hami_0 = -\hbar^2 \nabla^2/2m + q$ with $m$ being the atomic
mass.  The spin-mixing interactions are described through $\hamiint
= c_2 n(\mathbf{r})$ where $n(\mathbf{r})>0$ is the density of the
condensate and $c_2 = 4 \pi \hbar^2 (a_2 - a_0)/3 m$ relates to the
$s$-wave scattering lengths $a_f$ for binary collisions among atoms
with total spin $f \in \{0,2\}$ \cite{ho98,ohmi98}.  For a
ferromagnetic Bose gas, $c_2<0$.  Here, we restrict the spin
excitations to the volume of the condensate.

We find it convenient to consider separately the real and imaginary
parts of $\phi_a$, i.e.\ we consider the fields
$Z_a(\mathbf{r})=\phi^\dagger_a + \phi_a$ and $P_a(\mathbf{r}) = i
(\phi^\dagger_a - \phi_a)$, which commute as $[Z_a(\mathbf{r}),
P_a(\mathbf{r}^\prime)] = -2 i\,
\delta^3(\mathbf{r}-\mathbf{r}^\prime)$. Considering the linearized
equations of motion for these operators, obtained from $\hamispin$
given above, one identifies normal dynamical modes by defining a set
of functions $\zeta_n(\mathbf{r})$, spanning the condensate volume,
as solutions to the following differential equation:
\begin{equation}
\left( \hami_0 + 2 \hamiint \right) \zeta_n(\mathbf{r}) = E_n^2
\hami_0^{-1} \zeta_n(\mathbf{r}). \label{eq:zetadef}
\end{equation}
It follows that $E_n^2$ and $\zeta_n(\mathbf{r})$ are real and that
we may normalize the mode functions so that $\int d^3 \mathbf{r} \,
\zeta_n^* \hami_0^{-1} \zeta_m = \delta_{n,m} / 2$. With the aid of
these functions, we define the magnon-excitation mode operators
(omitting the polarization index)
\begin{eqnarray}
Z_n & = & \int d^3 \mathbf{r} \, \zeta_n(\mathbf{r}) \hami_0^{-1}
Z(\mathbf{r}) \label{eq:zmode}\\
P_n & = & \int d^3 \mathbf{r} \, \zeta_n(\mathbf{r}) P(\mathbf{r}).
\label{eq:pmode}
\end{eqnarray}
These operators are found to be canonically conjugate, satisfying
the equations of motion $\hbar \dot{Z}_n= P_n$ and $\hbar \dot{P}_n
= - E_n^2 Z_n$ and commutation relations $[ Z_n, P_m ] = - i
\delta_{n,m}$.  The canonical evolution of magnon excitation mode
$n$ is akin to that of a harmonic oscillator with frequency
$\omega_n = \sqrt{E_n^2}/\hbar$, with $E_n^2>0$ defining stable
magnon excitations and $E_n^2<0$ defining unstable modes.

The action of this canonical evolution on the initial fluctuations
in each normal mode may be discerned by examining the scaled
covariance matrix
\begin{eqnarray}
\mathcal{C}_{n}(t) & = & \mbox{Re} \left\langle \left(
\begin{array}{c} \tilde{Z}_n(t) \\ \tilde{P}_n(t) \end{array} \right) \left(
\begin{array}{cc} \tilde{Z}_n(t)  & \tilde{P}_n(t) \end{array} \right)
\right\rangle
\end{eqnarray}
where the mode operators are scaled by their initial rms
fluctuations.  It is important to note that this scaling
circularizes the initial fluctuations of the normal mode operators,
but that these initial fluctuations do not have the minimum
uncertainty. One finds, however, that the fluctuations do reach
their minimum product for excitations of a homogeneous condensate
and approximate the minimum product for inhomogeneous condensates
with densities varying only on long length scales.

The Hermitian covariance matrix may be expressed as
$\mathcal{C}_{n}(t) = \mathcal{R}_n(t) \mathcal{S}_n(t)
\mathcal{R}^\dagger_n(t)$ with $\mathcal{S}_n(t)$ being diagonal
with eigenvalues
\begin{equation}
S^{\pm} = 1 + 2 \kappa \sin^2 \omega_n t \pm 2 \sqrt{\kappa \sin^2
\omega_n t \left( 1 + \kappa \sin^2 \omega_n t\right)}
\label{eq:spmkappa}
\end{equation}
with $\kappa = (\chi + \chi^{-1} - 2)/4$ and $\chi = E_n^2 \langle
Z_n^2(0) \rangle/\langle P_n^2(0) \rangle$.  We note that $\kappa
\sin^2 \omega_n t$ is positive for both stable and unstable magnon
modes.

For stable magnon modes, the initial spin fluctuations rotate in the
$\tilde{Z}_n$-$\tilde{P}_n$ plane and also experience periodic,
bounded dilatations and contractions.  In contrast, for unstable
magnon modes, the spin fluctuations become progressively squeezed
along one quadrature axis and amplified along the other, with the
squeezed/amplified axes defined by the rotation matrix
$\mathcal{R}_n(t)$.  The fluctuations and dynamical evolution of
these mode operators relate directly to those of normalized
spatial-mode measurement observables given as
\begin{eqnarray}
\mtildeone_n = \int d^3 \mathbf{r} \, \frac{\hami_0^{-1}
\zeta_n(\mathbf{r}) }{n^{1/2}(\mathbf{r})
\langle Z_n^2(0) \rangle^{1/2}} \mone(\mathbf{r}) \label{eq:normmone}\\
\mtildetwo_n = \int d^3 \mathbf{r} \, \frac{ \zeta_n(\mathbf{r})
}{n^{1/2}(\mathbf{r}) \langle P_n^2(0) \rangle^{1/2}}
\mtwo(\mathbf{r}) \label{eq:normmtwo}
\end{eqnarray}
corresponding to $\tilde{Z}_n$ and $\tilde{P}_n$, respectively, at
$t=0$. Spatially resolved spin measurements tailored to detecting
the squeezed and amplified quadratures of the $n$th normal mode are
defined as time-dependent linear combinations of $\mtildeone_n$ and
$\mtildetwo_n$ through the rotation matrix $\mathcal{R}_n(t)$.

\subsection{Evolution of a homogeneous spinor condensate}
\label{sec:homo}

For a homogeneous condensate, the evolution following a quench of
the initially paramagnetic quantum gas is expressed simply.
Translational symmetry allows one to identify normal modes \emph{a
priori} as possessing Fourier components only with magnitude $k$.
The normal-mode equation (\ref{eq:zetadef}) is satisfied by the
functions $ \zeta_{\mathbf{k},\theta} = \sqrt{2/V} \,
\cos(\mathbf{k}\cdot\mathbf{r} + \theta)$,  with $V$ being the
condensate volume, and with the magnon-excitation spectrum given as
$E_k^2 = (\epsilon_k + q)(\epsilon_k + q - q_0)$ where $\epsilon_k =
\hbar^2 k^2/2 m$ \cite{lama07quench}.  Unstable modes exist for $q$
below a critical value $q_0 = - c_2 n$, specifically within the
range $|\epsilon + q - q_0/2| < q_0/2$.

The initial fluctuations of the normal mode operators defined with
respect to the mode functions $\{ \zeta_{\mathbf{k},\theta} \}$ are
given as $\langle Z_{\mathbf{k},\theta}^2 \rangle =
(\epsilon_k+q)^{-2}$ and $\langle P_{\mathbf{k},\theta}^2 \rangle
=1$, giving $\kappa = q_0^2/(4 E_k^2)$. The normalized covariance
matrix for modes with wave vector $k$ is given as
\begin{equation}
\mathcal{C}_k(t) = \left( \begin{array}{c c} 1 -
\frac{q_0}{\epsilon_k + q} \sin^2(\omega_k t) &  \frac{q_0}{\hbar \omega_k} \cos(\omega_k t) \sin(\omega_k t) \\
\frac{q_0}{\hbar \omega_k} \cos(\omega_k t) \sin(\omega_k t) & 1 +
\frac{q_0}{\epsilon_k + q - q_0} \sin^2(\omega_k t)
\end{array} \right).
\label{eq:covhomo}
\end{equation}

\begin{figure}[tb]
\centering
\includegraphics[width=0.6\textwidth]{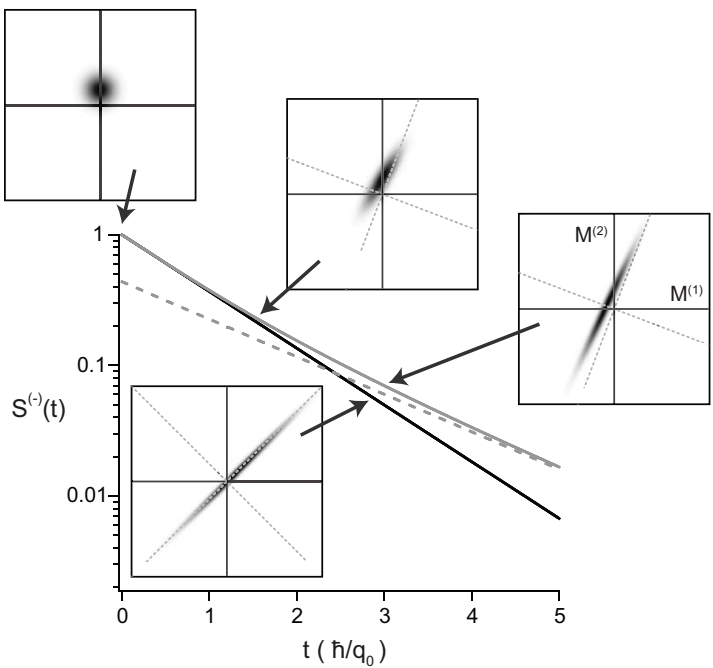}
\caption{Factor $S^{(-)}$ by which spin fluctuations are reduced
below the standard quantum limits, plotted vs.\ evolution time $t$
measured in units of $\hbar / q_0$.  Solid curves are shown for
momentum-space spin-fluctuation modes with either maximum gain,
obtained with $\epsilon_k + q = q_0/2$ (black), or lower gain, with
$\epsilon_k + q = 7 q_0/8$ (gray).  For the latter case, the
long-time limit for $S^{(-)}$ is shown as a dashed line.  Insets
show the corresponding spin fluctuations in the $M^{(1)}$-$M^{(2)}$
plane (the $Q$ representation) for the settings and times indicated
by arrows.  The long-time limits for the orientations of the
squeezing and amplification quadratures are indicated in the insets
by dashed gray lines.} \label{fig:sminus_figure}
\end{figure}

We see that the dynamically unstable momentum-space modes are
progressively squeezed over time (figure \ref{fig:sminus_figure}).
The maximally unstable mode, with $\omega_k^2 = -(q_0/2\hbar)^2$, is
accessed for $q < q_0/2$.  For this mode, the orientation of the
squeezing axis is fixed at an angle $\vartheta = - \pi/4$ in the
$M^{(1)}$-$M^{(2)}$ plane, and the greatest squeezing is achieved.
Away from this maximally squeezed mode, the orientation of the
squeezing/amplification axes varies in time.  At long times, defined
by $\kappa \sin^2(\omega_k t) = \frac{q_0^2}{4 \hbar^2 |\omega_k|^2}
\sinh^2(|\omega_k| t) \gg 1$, spin fluctuations at dynamically
unstable wave vectors are squeezed by a factor that decreases
exponentially in time and is given as
\begin{equation}
S^{(-)} \simeq \frac{4 \hbar^2 |\omega_k|^2}{q_0^2} e^{-2 |\omega_k|
t}. \label{eq:longtime}
\end{equation}
The orientation of the squeezed quadrature axis is given by the
relation $\tan \vartheta = - \hbar |\omega_k| / (\epsilon_k + q)$,
with $\vartheta$ subsequently varying between 0 and $- \pi/2$.

This momentum-space treatment describes how the dynamical
instabilities of a ferromagnetic $f=1$ spinor Bose-Einstein
condensate may be used as a mode-by-mode parametric amplifier, both
for the amplification of small spin fluctuations atop the polar
condensate, as explored in Refs.\
\cite{lesl09amp,klem09multi,sau09}, and also for the generation of a
spin-squeezed quantum field.  This field may be manipulated further
to suit a particular metrological application.  For example, spin
squeezing may first be prepared at a particular wave vector
$\mathbf{k}$ by allowing the system to evolve for a given time at $q
= q_1$ for which $\omega_k^2<0$.  Thereafter, the quadratic shift
may be adjusted to $q=q_2$ at which the spin modes at wave vector
$\mathbf{k}$ are dynamically stable.  Under such stable evolution,
the spin fluctuations undergo rotation in the $M^{(1)}$-$M^{(2)}$
plane, so that squeezing can be generated along any desired
quadrature axis.   More general control of the $k$-dependent
rotation angle could be realized by applying a $k$-dependent
optically induced quadratic Zeeman shift, e.g.\ by taking advantage
of the band structure of an appropriately tuned and polarized
optical lattice.

\section{Locality in spin squeezing and measurement}
\label{sec:locality}

It is illuminating to consider the evolution of a spatially extended
$f=1$ quantum field in position space. The spin-dependent contact
interactions to which we have ascribed the generation of a
correlated quantum field act locally; thus, we would expect spatial
correlations to propagate at a finite rate across the spinor
condensate.  Yet, the above description for the homogeneous system
(section \ref{sec:homo}) dealt with measurement observables that
incorporate correlations between particles at infinite range. In
contrast, given the local origin of spin correlations, we expect
that spin squeezing should be effectively observed by local
measurements of the collective atomic spin on a finite number of
atoms.

One approach to tracing the spatial evolution of spin correlations
is to consider correlations among the position-space
spin-fluctuation measurement densities, via the covariance matrix
defined as
\begin{equation}
\mathcal{C}(\mathbf{r}, \mathbf{r}^\prime; t) = \mbox{Re}
\left\langle \left(
\begin{array}{c} \Delta M^{(1)}_a(\mathbf{r}) \\ \Delta M^{(2)}_a(\mathbf{r}) \end{array} \right)_t  \left(
\begin{array}{cc} \Delta M^{(1)}_a(\mathbf{r}^\prime) & \Delta M^{(2)}_a(\mathbf{r}^\prime)\end{array} \right)_t
\right\rangle.
\end{equation}
For the uncorrelated condensate prepared at time $t=0$, we have
$\mathcal{C}(\mathbf{r},\mathbf{r}^\prime; 0) = \mathcal{I} \, n \,
\delta^3(\mathbf{r} - \mathbf{r}^\prime)$, i.e.\ correlations are
purely local.  At later times, for the initially homogeneous
condensate,
\begin{equation}
\mathcal{C}(\mathbf{r}, \mathbf{r}^\prime; t) = \frac{n}{(2 \pi)^3}
\int d^3 \mathbf{k} \, e^{i \mathbf{k} \cdot ( \mathbf{r} -
\mathbf{r}^\prime)} \mathcal{C}_k(t).
\end{equation}

Now, let us inquire as to the spatial range of spin-squeezing
produced by spin-dependent contact interactions by asking whether
finite-volume spatial-mode operators may be defined that will
capture the spin squeezing produced in a homogeneous spinor
condensate. We consider two possibilities.  First, we define
``quadrature-tuned'' measurement mode functions through their
Fourier transform as
\begin{equation}
\left( \begin{array}{c c} A(\mathbf{k}) & B(\mathbf{k}) \\
C(\mathbf{k}) & D(\mathbf{k}) \end{array} \right) = f(k)
\mathcal{R}_k^\dagger(t)
\end{equation}
where $f(k)$ is real. Now, following the evolution time $t$,
evaluating the covariance matrix with respect to these
spin-fluctuation observables, we obtain simply
\begin{equation}
\mathcal{C}_{qt}(t) = \frac{n V}{(2 \pi)^3} \, \int d^3 \mathbf{k}
\, f^2(k) \mathcal{S}_k(t).
\end{equation}
We consider also a second, simpler, ``uniform quadrature'' mode
function with Fourier transform
\begin{equation}
\left( \begin{array}{c c} A(\mathbf{k}) & B(\mathbf{k}) \\
C(\mathbf{k}) & D(\mathbf{k}) \end{array} \right) = f(k)
\mathcal{R}_{\kmax}^\dagger(t)
\end{equation}
where $\kmax$ is the wave vector for which maximal squeezing is
achieved.

Now we are in a position to consider forms of the measurement
operators appropriate for capturing squeezing achieved by the spinor
Bose-Einstein condensate amplifier.  We examine the long-time limit
(equation \ref{eq:longtime}), and consider amplification under two
conditions.

\subsection{Shallow quench}

Under the condition $q_0/2 \leq q < q_0$, the maximally unstable
mode occurs at zero wave vector. Expanding about $\mathbf{k} = 0$,
we approximate
\begin{equation}
|\omega_k| \simeq \frac{\sqrt{q(q_0 - q)}}{\hbar} +
\left(\frac{\hbar}{m} \right) \, \left(\frac{(q_0 - 2 q)}{4 \sqrt{q
(q_0 - q)}} \right) \, k^2 = \omega_{k = 0} - D \, k^2.
\end{equation}
We note that $D$, proportional to $\hbar / m$,  has the units of a
diffusion constant. Let us now define the measurement operator
through
\begin{equation}
f(k) = \frac{8 \pi^{3/2} \sigma^3}{\sqrt{V}}\, e^{- \sigma^2 k^2}
\end{equation}
so that the quantity $(A D - B C) > 0$ is a 3D Gaussian with peak
value of unity at the origin, and with a volume of $v = (2 \pi
\sigma^2)^{3/2}$.

Evaluating the correlation matrix, we now find fluctuations of the
$M^{(1)}_{A.B.C.D}$ operator are squeezed with respect to their
variance at $t=0$. An estimate for the squeezing achieved using the
quadrature-tuned mode function may be obtained analytically. If we
assume that $\sigma^2
> D t$, then the reduction in the variance is given approximately as
\begin{equation}
S^{(-)}_{qt} \simeq \frac{4 q (q_0 - q)}{q_0^2} \left(1 - \frac{ D
\, t}{\sigma^2}\right)^{-3/2} e^{-2 \omega_{k=0} t}.
\label{eq:shallowlimit}
\end{equation}
That is, the spin correlations produced in the quenched spinor
Bose-Einstein condensate are indeed local, contained within a volume
that grows diffusively in time.

\begin{figure}[tb]
\centering
\includegraphics[width=0.6\textwidth]{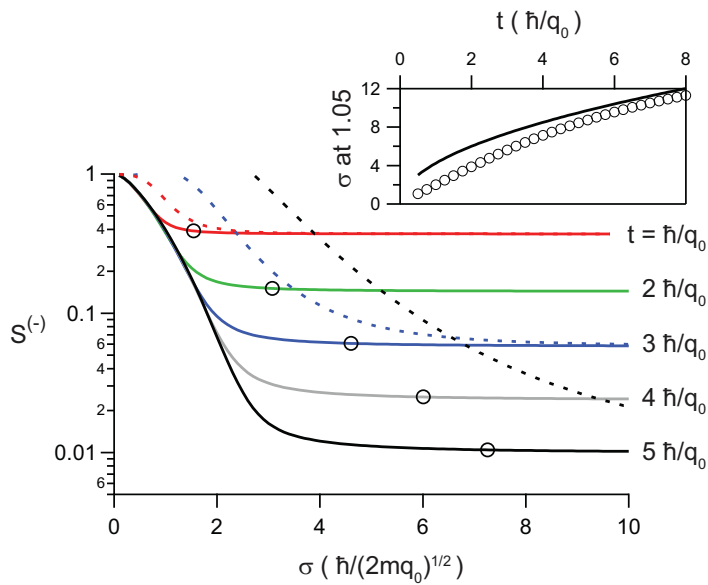}
\caption{Degree of squeezing, $S^{(-)}$, obtained by local
measurements after preparing an uncorrelated $|m_z = 0\rangle$
condensate and quenching it to a regime of dynamical instabilities
with $q/q_0 = 3/4$.  $S^{(-)}$ is calculated for variable $\sigma$
for quadrature-tuned measurement modes (solid lines) at evolution
times $t q_0 / \hbar = \{1,2,3,4,5\}$, and for uniform-quadrature
(dashed lines) at times $t q_0 / \hbar = \{1,3,5\}$.  The
quadrature-tuned measurement mode is found to capture the maximal
squeezing produced in the system within a measurement volume,
parameterized by $\sigma$, that grows only slowly with time.  Open
circles highlight values of $\sigma$ at which $S^{(-)}$ is within
5\% of the single-mode $k=0$ value.  As shown in the inset, the
dimension $\sigma$ at which this condition is achieved (open
circles) is approximated at late times $t$ according to  equation
\ref{eq:shallowlimit} (solid line).  In contrast, uniform-quadrature
measurements must be made on larger volumes to observe the maximum
attainable squeezing.} \label{fig:shallowsspin}
\end{figure}

These approximations are confirmed by numerical calculations,
results of which are shown in figure \ref{fig:shallowsspin}.  The
quadrature-tuned spin-fluctuation measurement is indeed effective in
capturing the spin squeezing produced in this system within a
measurement volume that grows only slowly with evolution time.  In
contrast, a uniform-quadrature measurement requires larger volumes
to capture this squeezing.  This requirement is explained by the
finite differences between the quadrature axes along which squeezing
is achieved at different wave vectors.  The higher the degree of
squeezing one wishes to detect, the more precisely must the
measurement quadrature axes be tuned to the spin-squeezed axis by
restriction to a narrow range of wave vectors.

\subsection{Deep quench}
For a deep quench, with $q < q_0/2$ the maximum instability occurs
on a sphere in $k$-space, with radius $\kmax$ defined by $\epsilon_k
+ q = q_0/2$. About this shell, the gain is given approximately as
\begin{equation}
|\omega_k| \simeq \frac{q_0}{2 \hbar} - \frac{\hbar}{m} \frac{q_0 -
2 q}{q_0} (k - \kmax)^2 = \frac{q_0}{2 \hbar} - D (k - \kmax)^2.
\end{equation}
To capture this squeezing, we define a measurement mode through the
following relation:
\begin{equation}
f(k) = X \, e^{- \sigma^2 (k - \kmax)^2} \label{eq:fdeep}
\end{equation}
with $X$ chosen so that $A D - B C= 1$ at $\mathbf{r}=0$. For large
$\sigma$, the volume defined by this function tends toward $v =
\sqrt{2 \pi^3} \, \sigma / \kmax^2$, so that the effective radius of
the measurement volume is $r \sim (\sigma / \kmax^2)^{1/3}$.

\begin{figure}[tb]
\centering
\includegraphics[width=0.6\textwidth]{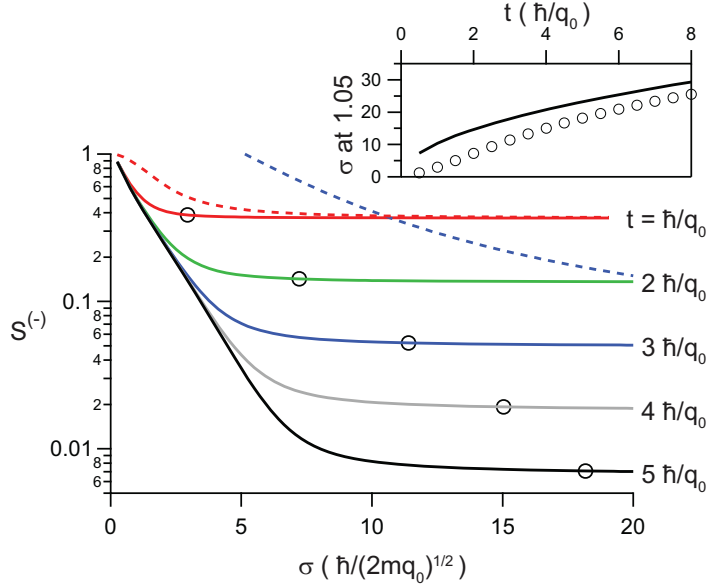}
\caption{Degree of squeezing, $S^{(-)}$, obtained by local
measurements following a deep quench to $q = -2 q_0$.
Quadrature-tuned measurements taken at evolution times $t q_0 /
\hbar = \{1,2,3,4,5\}$ evince the maximal single-mode spin squeezing
for mode functions given by equation \ref{eq:fdeep} with $\sigma$
growing only slowly with time (solid lines). Open circles highlight
values of $\sigma$ at which $S^{(-)}$ is within 5\% of the
single-mode $k=\kmax$ value.  As shown in the inset, the dimension
$\sigma$ at which this condition is achieved (open circles) is
approximated at late times $t$ according to  equation
\ref{eq:deeplimit} (solid line).  Uniform-quadrature mode functions,
shown here for $t q_0/\hbar = \{1, 3\}$, require much larger
$\sigma$ to capture the full spin squeezing in this system.}
\label{fig:deepsspin}
\end{figure}

Now we determine the degree of spin squeezing captured by this
measurement operator.  Repeating calculations as above, with the
requirement $\sigma^2 > D t$, we have
\begin{equation}
S^{(-)}_{qt} \simeq \left(1 - \frac{D t}{\sigma^2}\right)^{-1/2}
e^{- q_0 t / \hbar}. \label{eq:deeplimit}
\end{equation}
Again, we see that local measurements will capture the spin
squeezing produced by the spinor Bose-Einstein condensate amplifier.
Here, the correlation volume grows slower than diffusion, with
radius scaling only as $r \propto t^{1/6}$.  Numerical evaluations
of $S^{(-)}$ are presented in figure \ref{fig:deepsspin}, confirming
the rapid convergence onto the single-mode maximal degree of
squeezing for finite-volume, quadrature-tuned measurement modes.

\section{Experimental considerations}

We conclude with a discussion of experimental prospects for
realizing and detecting multi-mode spin squeezing in a quenched
ferromagnetic $f=1$ condensate.  We focus on the case of $^{87}$Rb,
which has been shown experimentally
\cite{schm04,chan04,wide06precision} and theoretically
\cite{klau01rbspin,vankemp02} to be ferromagnetic, with $\Delta a =
-1.4(3) \, a_B$ with $a_B$ being the Bohr radius.  The behaviour of
condensates prepared in the  $|m_z = 0\rangle$ state and quenched to
the regime of dynamic instabilities by a rapid change of $q$ have
been studied using \emph{in situ} magnetization-sensitive imaging.
Pertinent to the present discussion, the amplification of spin
fluctuations following quenches to different values of $q$ was
measured precisely in Ref.\ \cite{lesl09amp}, and such amplification
was found to be roughly consistent with quantum-limited
amplification of quantum fluctuations in the initial state, with a
gain, measured by the increase in the variance of spin fluctuations,
as high as +30 dB.

In contrast to previous studies on the parametric amplification of
spin fluctuations \cite{lesl09amp,sau09}, here we have considered
the squeezing that may accompany such amplification.  We assume
experimental conditions similar to those of Refs.\
\cite{sadl06symm,lesl09amp}, i.e.\ an optically trapped $f=1$
$^{87}$Rb condensate of $N = 2.0 \times 10^6$ atoms with peak
density $n = 2.6 \times 10^{14} \, \mbox{cm}^{-3}$. At this density,
the characteristic time scale for spin mixing in the spinor
condensate is $\hbar / q_0 = 8\, \mbox{ms}$ and the characteristic
length scale is $(\hbar^2 / (2 m q_0))^{1/2} = 1.8 \, \mu\mbox{m}$.
Let us estimate the maximum observable degree of spin squeezing,
taking into account non-linearities, atom loss, and the spatial
inhomogeneity of the condensate.

\subsection{Single-mode nonlinearities and atom loss}
\label{sec:atomloss}

Thus far, we have treated spin dynamics in a quenched $f=1$ spinor
condensate exclusively through the Bogoliubov approximation
(equation \ref{eq:hspinbogo}), i.e.\ neglecting high-order terms in
the spin fluctuation operators and also the depletion of the
$|m_z=0\rangle$ condensate due to spin mixing.  Within this linear,
undepleted-pump approximation, we obtain the unphysical result,
expressed in equation \ref{eq:spmkappa}, that the minimum
spin-quadrature variance $S^{(-)}$ in each unstable spin mode tends
exponentially to zero. As in quantum-optical systems, sensible
limits to squeezing are recovered by going beyond the linearized
equations of motion and the undepleted-pump approximation.

These limits are partly elucidated in the full dynamics of a single
spin mode. We consider therefore the dynamics of a ferromagnetic
spinor condensate in the single-mode approximation.  Such a
treatment would be appropriate for a spinor condensate spanning a
volume $v \lesssim (\hbar^2 / 2 m q_0)^{3/2}$, realized
experimentally with an atom number of $n v \lesssim 1000$.  The full
Hamiltonian is now given as
\begin{equation}
\hami \! = \! \sum_{a\in\{x,y\}}\! \left\{\left[-\frac{c_2}{2 v}
\left(\left.\phi^\dagger_a\right.^2 \phi_z^2 +
\left.\phi^\dagger_z\right.^2 \phi_a^2 \right) + q \phi^\dagger_a
\phi_a\right] +  \frac{c_2}{v} \phi^\dagger_z \phi_z \phi^\dagger_a
\phi_a \right\} - \frac{c_2}{2 v} \left(\phi^\dagger_x \phi_y -
\phi^\dagger_y \phi_x \right)^2
\end{equation}
where the Bose operators $\phi_e$ annihilate particles in polar
state $|\phi_e\rangle$ and in the single spatial mode of the
condensate.

In this expression, the single-polarization term $\propto
\left(\left.\phi^\dagger_a\right.^2 \phi_z^2 +
\left.\phi^\dagger_z\right.^2 \phi_a^2 \right)$ matches the two-axis
counter-twisting Hamiltonian described in Ref.\ \cite{kita93}.  On
its own, this number-conserving Hamiltonian produces squeezing that
saturates with time due to depletion of the pump (the population
$\phi^\dagger_z \phi_z$), reaching the Heisenberg limit with
$S^{(-)} \propto 1/N$ \cite{andr02corr}.  However, the Hamiltonian
contains additional terms: the externally imposed and pump-dependent
quadratic Zeeman shifts and inter-polarization coupling.  These
nonlinear terms restrict spin squeezing to sub-Heisenberg scaling,
with $S^{(-)} \propto 1/\sqrt{N}$. Nevertheless, the expected
single-mode squeezing is significant.  We have performed exact
single-mode calculations for a 1000-atom sample, and find a
reduction of spin fluctuations by as much as 20 dB.

The single-mode limit allows us to discuss an additional limitation
arising from the decay of atoms from the trap.  Atom loss can be
modeled by adding a noise source and a density-dependent loss rate
in the coherent evolution of the Bose fields \cite{andr02corr}. We
have performed numerical calculations including such terms,
representing quantum noise in the evolution of a classical field as
in the truncated Wigner approximation (TWA) \cite{gard00bookqnoise}.
In accord with experimental observations, we assume an initial
per-atom loss rate of 1 $\mbox{s}^{-1}$.  We find squeezing in a
$N=1000$ atom sample can still be the level of -20 dB, a result that
is not surprising given the very short spin-mixing time as compared
to the lifetime of trapped atoms.

\subsection{Effects of inhomogeneity and multi-mode nonlinearities}

Experimental realizations of spinor Bose-Einstein condensates also
differ from our idealized treatment in that the condensates are
inhomogeneous, and, thus, the adequacy of a linearized normal-mode
description to long-time dynamical evolution is less clear than in
the homogeneous case.  To assess the influence of mixing between
normal modes and also atom losses for this situation, we have
performed numerical simulations of the evolution of a
quasi-one-dimensional quenched spinor Bose-Einstein condensate. Such
a condensate is assumed sufficiently tightly confined in two radial
dimensions that the spinor wavefunction is radially uniform, while
being more weakly confined in the third dimension, along which the
wavefunction is allowed to vary.  Specifically, we consider a
condensate with Thomas-Fermi radii of 3 and 200 $\mu\mbox{m}$ along
the radial and axial dimensions, respectively. The central, radially
averaged condensate density $n_0$ determines the spin mixing
interaction strength to be $2 |c_2| n_0 = h \times 15 \, \mbox{Hz}$.

The quantum evolution of this spinor condensate is calculated using
the TWA to simulate the effects of quantum noise and squeezing
\cite{sau09}. This approximation involves adding random noise to the
initial mean-field condensate wavefunction, with the noise variance
determined, in this case, by the one-dimensional condensate density.
Temporal evolution is thereafter considered according to classical
wave equations, which include the full non-linear dynamics given by
spin-mixing Hamiltonian (the non-linear Schr\"{o}dinger equation)
and also the effects of atom loss (through additional noise and loss
terms, described above). Expectation values, variances, and
correlations for various measurement operators are quantified by
repeating the calculations for many instances of the initial noise.
These calculations were found convergent at a temporal resolution of
$3.5$ $\mu$s and spatial resolution of $0.5$ $\mu$m.

\begin{figure}[tb]
\centering
\includegraphics[width=\textwidth]{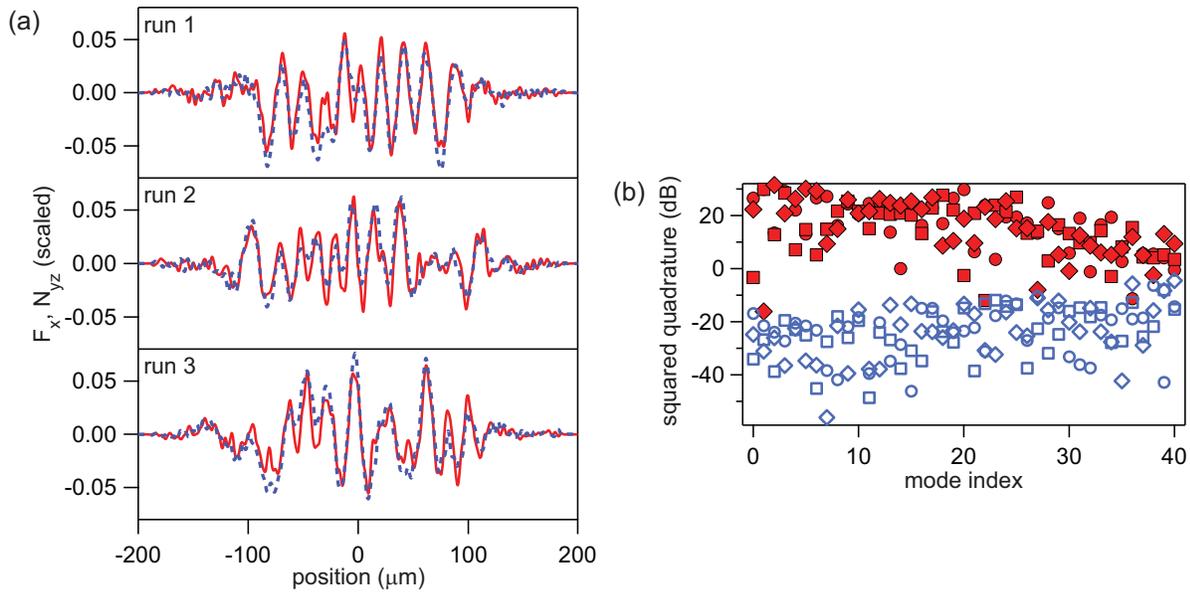}
\caption{Numerical simulations of amplification and squeezing of
spin fluctuations in a quenched, trapped spinor Bose-Einstein
condensate.  The evolution of quantum noise is simulated via the
TWA.  (a) The results of simulation runs, each with different
instances of random initial noise, are shown at $t=65$ ms after a
quench to the regime of dynamical instabilities (see text for
details).  One polarization of local spin fluctuation observables
(components of magnetization (red solid line) and nematicity (blue
dashed line)) is shown, scaled to unity for a fully spin polarized
condensate at its center.  Strong local correlation of the amplified
spin fluctuations is evident.  (b) Squared values of the amplified
(solid symbols) and squeezed (closed symbols) normal-mode quadrature
measurement operators, evaluated for runs 1 (circles), 2 (squares)
and 3 (diamonds), scaled to their initial statistical variance,
indicate a reduction of initial spin fluctuations by 20 to 30 dB.}
\label{fig:threeruns}
\end{figure}

We present results of a numerical simulation of an initially polar,
trapped, $^{87}$Rb $f=1$ spinor condensate quenched at $t=0$ to a
quadratic Zeeman shift of $q = h \times 2.5 \, \mbox{Hz}$.  The
results for several numerical runs, i.e.\ for three random instances
of initial noise as evaluated according to the TWA, are presented in
figure \ref{fig:threeruns}a. Dynamical instabilities following the
quench to low $q$ result in amplified fluctuations in the transverse
components of the magnetization and the related components of the
nematicity; only one polarization of the observables, $F_x$ and
$N_{yz}$, is shown in the figure. The strong local correlation
between these observables is clearly evident, reflecting the
$\vartheta \simeq \pi/4$ angle of the amplified quadrature axis for
the maximally unstable normal modes.

\begin{figure}[tb]
\centering
\includegraphics[width=0.6\textwidth]{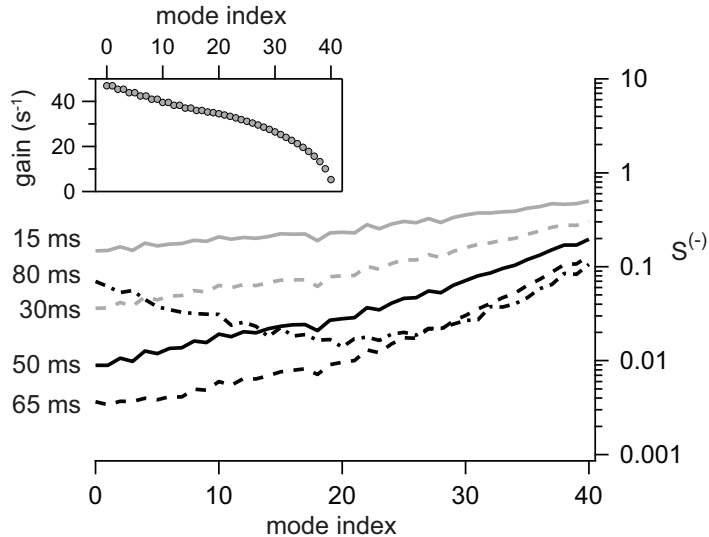}
\caption{Suppression of spin fluctuations in a quenched, trapped
spinor Bose-Einstein condensate, evaluated for normal dynamical
modes (labeled by mode index $n$) at several times of evolution
(labeled on left of graph). The calculated early-time gain of these
unstable modes is shown in the inset.  The measurement variance in
the squeezed quadrature mode $S^{(-)}$ diminishes exponentially in
time, saturating at an evolution time of $t = 65$ ms to a minimum
value of $3.6 \times 10^{-3}$ for the highest-gain mode, and then
growing due to saturation and nonlinear effects.} \label{fig:intrap}
\end{figure}

Less evident in these position-space plots is the sharp reduction in
the initial fluctuations of the normal-mode measurement observables.
To reveal this reduction, we identify the normal dynamical modes
based on the initial condensate distribution. The temporal gain
determined for the first 41 unstable modes is shown in the inset of
figure \ref{fig:intrap}, sorted in decreasing order of temporal gain
($|\omega_n|$) and labeled by an integer mode index. Normalized
spatial-mode measurement operators are then formed as defined in
equations \ref{eq:normmone} and \ref{eq:normmtwo}, and the values of
these operators at different evolution times $t$ are evaluated for
an ensemble of 600 numerical runs. From these values, we evaluate
the scaled covariance matrices $\mathcal{C}_n$ numerically and
determine therefrom the quadrature axes that are amplified and
squeezed, and also the measurement variance along those axes.

The squares of the measurement outcomes along the amplified and
squeezed quadrature axes for the three simulation runs discussed
above are shown in figure \ref{fig:threeruns}b.  Aside from
pronounced variations according to the initial random values
assigned to the quantum noise in these simulations, these data
indicate a suppression of spin fluctuations between 20 and 30 dB.

The ensemble averaged reduction $S^{(-)}$ in the initial spin
fluctuations, determined for each mode $n$ and at different
evolution times, is shown in figure \ref{fig:intrap}.  We find the
temporal evolution of the dynamically unstable modes causes the
measurement variance of the squeezed quadrature operators to
diminish exponentially in time for evolution times up to around 60
ms, after which the variance saturates and then begins growing. The
minimum variance of around $3.6 \times 10^{-3}$ or -24 dB is
achieved for the unstable modes with highest gain (lowest mode
index) at $t = 65 \, \mbox{ms}$.  This maximum degree of squeezing
matches well with the estimated effects of nonlinearities and atom
loss presented above.

\section{Conclusion}

We have presented a formalism for the evaluation of spin squeezing
in ensembles of high-spin atoms. This formalism identifies several
independent planes of spin fluctuations atop an arbitrary coherent
spin state, revealing a greater resource for applications of spin
squeezing in quantum information and metrology than for ensembles of
spin-1/2 particles.  Moreover, we developed an understanding of spin
squeezing in spatially extended quantum fields, where squeezing in
many spatial modes can be separately generated and measured. This
formalism is applicable to many contemporary experiments on spin
squeezing of atomic gases or of other distributed quantum objects.

Applying this formalism, we consider a source of spin squeezing in
an extended spinor Bose-Einstein condensate.  Referring to recent
experiments studying the evolution of $^{87}$Rb condensates quenched
to a regime of dynamical instabilities
\cite{sadl06symm,lesl09amp,klem09multi}, we show that the strong
amplification of spin fluctuations demonstrated in those experiments
is accompanied also by spin squeezing, i.e.\ that the dynamically
unstable normal modes act as quantum parametric amplifiers of spin
fluctuations. Treating experimental realities such as atom losses
and effects of confinement, we apply analytic and numerical tools to
determine that measurable spin squeezing better than 20 dB may be
achieved in such systems.

This treatment is relevant also to the question of symmetry breaking
in a quantum many-body system.  An initially polar-state spinor
Bose-Einstein condensate preserves the $SO(2)$ rotational symmetry
about the alignment axis.  The quantum quench considered in this
work, a rapid reduction in the quadratic Zeeman shift to induce
dynamical instabilities and parametric amplification of spin
fluctuations, traverses the atomic system across a symmetry-breaking
quantum phase transition, where the spin-dependent contact
interactions favor broken symmetry states of maximum magnetization
\cite{sadl06symm}.  In a closed noiseless quantum system, bona fide
symmetry breaking does not occur; rather, the system evolves
coherently to a symmetric superposition of broken-symmetry states.
The spin-squeezed states described in this work indeed represent
such a coherent superposition.

The metrological significance of spin squeezed quantum fields, and
also the question of symmetry breaking in quantum systems, provide
strong motivation for the detection and characterization of
squeezing in quenched spinor Bose gases.  For this purpose,
spatially resolved measurements of the magnetization of optically
trapped condensates have been demonstrated \cite{veng07mag}, and
optical means of measuring components of both the magnetization and
the nematicity have been suggested \cite{caru04imag}.  However,
sensitivity below the atomic shot noise level, required for the
direct detection of squeezing, is experimentally challenging and was
not achieved in those measurements.

Alternately, we propose that one could use the condensate itself as
a pre-measurement amplifier.  Following the generation of
inhomogeneous spin squeezing with the quadratic shift at $q_1$, one
may quickly increase the quadratic Zeeman shift.  The previously
unstable dynamical modes now evolve stably as the squeezed and
amplified spin-fluctuation quadratures rotate in their quadrature
spaces. After their rotation, the spin fluctuations may be again
parametrically amplified by returning the quadratic shift to $q_1$.
Spin squeezing produced by the first stage of amplification would
result in a reduction of spin fluctuations observed after the second
amplification stage, as compared to a single-stage amplification of
vacuum spin noise.  We have performed TWA-based numerical
simulations which confirm that this technique is applicable to
trapped spinor Bose-Einstein condensates at experimentally
accessible settings.

This work was supported by the NSF and the Army Research Office with
funding from the DARPA OLE program. Partial personnel and equipment
support was provided by the LDRD Program of LBNL under the Dept.\ of
Energy Contract No.\ DE-AC02-05CH11231. D.M.S.-K.\ acknowledges
support of the Miller Institute for Basic Research in Science, and
J.S. acknowledges the JQI-NSF-PFC for support.

\section*{References}

\begin{thebibliography}{10}

\bibitem{wine92squeeze}
D.J. Wineland et~al.
\newblock Spin squeezing and reduced quantum noise in spectroscopy.
\newblock {\em Phys. Rev. A}, 46:R6797, 1992.

\bibitem{kita93}
M.~Kitagawa and M.~Ueda.
\newblock Squeezed spin states.
\newblock {\em Phys. Rev. A}, 47(6):5138--5143, 1993.

\bibitem{schl10states}
M.~H. Schleier-Smith, I.~D. Leroux, and V.~Vuletic.
\newblock States of an ensemble of two-level atoms with reduced quantum
  uncertainty.
\newblock {\em Phys. Rev. Lett.}, 104(7):073604, 2010.

\bibitem{appe09}
J.~Appel et~al.
\newblock Mesoscopic atomic entanglement for precision measurements beyond the
  standard quantum limit.
\newblock {\em Proc. Natl. Acad. Sci. USA}, 106(27):10960--10965, 2009.

\bibitem{este08squeeze}
J.~Esteve et~al.
\newblock Squeezing and entanglement in a Bose-Einstein condensate.
\newblock {\em Nature}, 455(7217):1216--1219, 2008.

\bibitem{hald99}
J.~Hald, J.L. S\/{o}rensen, C.~Schori, and E.S. Polzik.
\newblock Spin squeezed atoms: a macroscopic entangled ensemble created by
  light.
\newblock {\em Phys. Rev. Lett.}, 83(7):1319, 2000.

\bibitem{kupr05}
D.~V. Kupriyanov et~al.
\newblock Multimode entanglement of light and atomic ensembles via off-resonant
  coherent forward scattering.
\newblock {\em Phys. Rev. A}, 71(3):032348, 2005.

\bibitem{stoc06tensor}
J.M. Geremia, J.K. Stockton, and H.~Mabuchi.
\newblock Tensor polarizability and dispersive quantum measurement of
  multilevel atoms.
\newblock {\em Phys. Rev. A}, 73(042112), 2006.

\bibitem{yash03higher}
V.~V. Yashchuk et~al.
\newblock Selective addressing of high-rank atomic polarization moments.
\newblock {\em Phys. Rev. Lett.}, 90(25):253001, 2003.

\bibitem{acos08hexa}
V.~M. Acosta et~al.
\newblock Production and detection of atomic hexadecapole at Earth's magnetic
  field.
\newblock {\em Optics Express}, 16(15):11423, 2008.

\bibitem{sadl06symm}
L.E. Sadler et~al.
\newblock Spontaneous symmetry breaking in a quenched ferromagnetic spinor Bose
  condensate.
\newblock {\em Nature}, 443:312, 2006.

\bibitem{lesl09amp}
S.R. Leslie et~al.
\newblock Amplification of fluctuations in a spinor Bose Einstein condensate.
\newblock {\em Phys. Rev. A}, 79:043631, 2009.

\bibitem{klem09multi}
C.~Klempt et~al.
\newblock Multiresonant spinor dynamics in a Bose-Einstein condensate.
\newblock {\em Phys. Rev. Lett.}, 103(19):195302, 2009.

\bibitem{veng07mag}
M.~Vengalattore et~al.
\newblock High-resolution magnetometry with a spinor Bose-Einstein condensate.
\newblock {\em Phys. Rev. Lett.}, 98(20):200801, 2007.

\bibitem{note:polar}
Specifically, we use the basis $|\phi_x\rangle = \left( |m_z =
1\rangle - |m_z
  = -1\rangle \right)/\sqrt{2}$, $\phi_y\rangle = i \left( |m_z = 1\rangle +
  |m_z = -1\rangle \right)/\sqrt{2}$, and $|\phi_z\rangle = |m_z = 0\rangle$.
  With this definition, the vector spin operator takes the form
  $\left(F_a\right)_{bc} = -i \epsilon_{abc} |\phi_a\rangle \langle \phi_b|$
  where the indices run over the Cartesian coordinates.

\bibitem{note:quadrupole}
We adopt the definition $N_{ab} = (F_a F_b + F_b F_a) - 4
\delta_{ab} / 3$
  where $a,b \in \{x,y,z\}$ and $F_a$ denotes an angular momentum component
  operator in matrix form.

\bibitem{gior03}
Paolo Giorda, Paolo Zanardi, and Seth Lloyd.
\newblock Universal quantum control in irreducible state-space sectors:
  Application to bosonic and spin-boson systems.
\newblock {\em Phys. Rev. A}, 68(6):062320, 2003.

\bibitem{note:necessary}
We emphasize that Bose-Einstein condensation, or even Bose-Einstein
statistics,
  are not necessary for the existence of spin squeezing in an extended,
  multi-mode quantum field. Much of the notation developed in this work may be
  extended also to squeezing with more generic constituents. Here, we focus on
  spinor Bose-Einstein condensates not only for notational convenience, but
  also to describe the role of spin-dependent interactions in creating such
  squeezing.

\bibitem{caru04imag}
Iacopo Carusotto and Erich~J. Mueller.
\newblock Imaging of spinor gases.
\newblock {\em J. Phys. B}, 37:S115, 2004.

\bibitem{lama07quench}
A.~Lamacraft.
\newblock Quantum quenches in a spinor condensate.
\newblock {\em Phys. Rev. Lett.}, 98(16):160404, 2007.

\bibitem{mias08}
G.~I. Mias, N.~R. Cooper, and S.~M. Girvin.
\newblock Quantum noise, scaling, and domain formation in a spinor
  Bose-Einstein condensate.
\newblock {\em Phys. Rev. A}, 77(2):023616, 2008.

\bibitem{ho98}
T.-L. Ho.
\newblock Spinor Bose condensates in optical traps.
\newblock {\em Phys. Rev. Lett.}, 81:742, 1998.

\bibitem{ohmi98}
T.~Ohmi and K.~Machida.
\newblock Bose-Einstein condensation with internal degrees of freedom in alkali
  atom gases.
\newblock {\em J. Phys. Soc. Jpn.}, 67:1822, 1998.

\bibitem{sau09}
J.D. Sau, S.R. Leslie, M.L. Cohen, and D.M. Stamper-Kurn.
\newblock Theory of domain formation in inhomogeneous ferromagnetic dipolar
  condensates within the truncated Wigner approximation.
\newblock {\em Phys. Rev. A}, 80:023622, 2009.

\bibitem{schm04}
H.~Schmaljohann et~al.
\newblock Dynamics of F = 2 spinor Bose-Einstein condensates.
\newblock {\em Phys. Rev. Lett.}, 92:040402, 2004.

\bibitem{chan04}
M.-S. Chang et~al.
\newblock Observation of spinor dynamics in optically trapped Rb Bose-Einstein
  condensates.
\newblock {\em Phys. Rev. Lett.}, 92:140403, 2004.

\bibitem{wide06precision}
A.~Widera et~al.
\newblock Precision measurement of spin-dependent interaction strengths for
  spin-1 and spin-2 Rb-87 atoms.
\newblock {\em New Journal of Physics}, 8:152, 2006.

\bibitem{klau01rbspin}
N.N. Klausen, J.L. Bohn, and C.H. Greene.
\newblock Nature of spinor Bose-Einstein condensates in rubidium.
\newblock {\em Phys. Rev. A}, 64:053602, 2001.

\bibitem{vankemp02}
E.G.M.~van Kempen, S.J.J.M.F. Kokkelmans, D.J. Heinzen, and B.J.
Verhaar.
\newblock Interisotope Determination of Ultracold Rubidium Interactions from
  Three High-Precision Experiments.
\newblock {\em Phys. Rev. Lett.}, 88:093201, 2002.

\bibitem{andr02corr}
A.~Andr\'{e} and M.D. Lukin.
\newblock Atom correlations and spin squeezing near the Heisenberg limit:
  Finite-size effect and decoherence.
\newblock {\em Phys. Rev. A}, 65:053819, 2002.

\bibitem{gard00bookqnoise}
C.W. Gardiner and P.~Zoller.
\newblock {\em Quantum Noise}.
\newblock Springer-Verlag, Berlin, Heidelberg, 2000.

\end{thebibliography}

\end{document}